\documentclass[copyright,creativecommons]{eptcs}
\providecommand{\event}{FTSCS 2012} 
\usepackage{breakurl}             

\newtheorem{definedef}{Definition}
\usepackage[pdftex]{graphicx}

\title{Property-based Code Slicing for Efficient Verification of OSEK/VDX Operating Systems\footnote{This work was partially supported by the Engineering Research Center of Excellence Program of Korea Ministry of Education, Science and Technology(MEST)/National Research Foundataion of Korea(NRF) (Grant 2012-0000473)
and the National Research Foundation of Korea Grant funded by Korean Government (2012R1A1A4A01011788).
 }}
\author{Mingyu Park  \qquad\qquad Taejoon Byun  \qquad\qquad Yunja Choi
\institute{School of Computer Science and Engineering\\
Kyungpook National University \\
Deagu, Korea}
\email{\quad  pqrk8805@gmail.com \quad\quad bntejn@gmail.com \quad\quad yuchoi76@knu.ac.kr\footnote{correspondence}}
}
\def\titlerunning{Property-based Code Slicing}
\def\authorrunning{ Mingyu Park \& Taejoon Byun \& Yunja Choi}
\begin{document}
\maketitle

\sloppy

\begin{abstract}
Testing is a de-facto verification technique in industry, but insufficient for identifying subtle issues due to its optimistic incompleteness.
On the other hand, model checking is a powerful technique that supports comprehensiveness, and is thus suitable for the verification of safety-critical systems.
However, it generally requires more knowledge and cost more than testing.
This work attempts to take advantage of both techniques to achieve integrated and efficient verification of OSEK/VDX-based automotive operating systems.
We propose property-based environment generation and model extraction techniques using static code analysis, which can be applied to both model checking and testing.
The technique is automated and applied to an OSEK/VDX-based automotive operating system, Trampoline. Comparative experiments
 using random testing and model checking for the verification of assertions in the Trampoline kernel code show how our environment generation and abstraction approach can be utilized for efficient fault-detection.

\end{abstract}

\section{Introduction}
The operating system is the core part of automotive control software; any malfunction
  can cause critical errors in the automotive system, which in turn may result in loss of lives and assets.
Testing has been widely used as a systematic and cost-effective safety analysis/assurance method~\cite{Broy:2006,Moessinger:IEEEsoftware10}, but its
optimistic incompleteness often misses critical problems and cannot guarantee the ``absence of wrong behavior".
As an alternative and complimentary technique, model checking~\cite{ModelCheckingBook,SPINBOOK} has been drawing attention from both academia and industry.

Model checking is a comprehensive formal verification technique, suitable for functional safety analysis. It can effectively identify subtle issues, such as
process dead lock, illegal behavior, and starvation, but may require more resources and domain knowledge. In particular, the use of model checking faces the following
challenges:

\begin{enumerate}
\item The size of model/code to be verified needs to be minimized to avoid state-space explosion.
\item Modeling of the environment, such as user tasks and hardware environment, is necessary and critical for embedded software.
\end{enumerate}

Since an operating system is a reactive system responding to environmental stimuli, the correctness of
 its behavior needs to be analyzed with respect to the behavior of its environments. A non-deterministic environment is typically used to over-approximate
 actual behavior, but it is often too expensive in model checking.
 The difficulty and importance of defining
 a {\it good} environment model has been addressed in a number of previous works~\cite{Penix:FMSD05,TDP03,GP02,Yatake:SPIN10,IS02,HP93}.

We note that these two problems apply to both model checking and testing. Though the level of comprehensiveness differs, both techniques rely on automated search techniques that are initiated by environmental stimuli. This is called environment model in model checking and test scenario in testing.
This work anticipates that the efficiency of automated verification techniques depends on the modeling of the environment and proposes
an application of property-based code slicing~\cite{Weiser84} for automatically generating an environment model using the data/function dependency analyzed from the operating system kernels.
The goal is to construct a valid and comprehensive usage model of the operating system with minimal dependency on the kernel code.

Our approach extracts functions that have a direct dependency on a given property to be verified and generates non-deterministic function-call sequences by imposing (1) external constraints from the OSEK/VDX standard~\cite{OSEK/VDX:Portal} for automotive operating systems, and (2) internal constraints identified from the function call structure of the operating system kernel. The external constraints are manually identified from the specifications of the standard and are imposed on the initially random sequence of function calls. The internal constraints are imposed by identifying the top-level functions using backward slicing from a given property and by computing the cone-of-influence from each top-level function using forward slicing.
The Environment model is defined as an arbitrary sequence of calls of those extracted functions. The operating system kernel is also abstracted as a collection of
extracted functions and its relevant code required for pre-processing them. This procedure reduces the size of the verification target and minimizes the behavior of the environment model.
The extraction and model construction process is automated with the aid of the static analysis tool Understand~\cite{Understand}.

The approach and the tool are applied to the verification of safety properties of the Trampoline operating system~\cite{Trampoline:Portal}, which is an open source automotive operating system compliant with OSEK/VDX.
Environment models are generated using the assertions identified from the kernel code, and the kernel code itself is reduced by including only those extracted functions and their relevant code.
The environment model is used to model-check/test the abstract code using CBMC~\cite{CBMC:TACAS04} and random testing. We compare their fault-detection capability, their comprehensiveness in terms of code coverage, and their efficiency in terms of resource consumption.

The remainder of this paper is organized as follows. Section~\ref{sec:related}  briefly discusses related work and Section~\ref{sec:background} provides the motivation for our work. Section~\ref{sec:overview} provides an overview of our approach and
 Section~\ref{sec:env_gen}  presents the methods and the process for the automated environment generation technique. Section~\ref{sec:env_setting} explains the environment settings for the collaborative
  verification, followed by experimental results and the evaluation using Trampoline OS as a case example in Section~\ref{sec:experiment}.
We conclude in Section~\ref{sec:conclusion}.

\section{Related Work}
\label{sec:related}

Environment modeling for efficient model checking has been an active research issue~\cite{Penix:FMSD05,TDP03,GP02,Yatake:SPIN10,IS02,HP93}.
Reference~\cite{HP93} is one of the earliest works concerning environment assumptions in verification. It introduced the $observer$ concept to represent assumptions about the environment. The approaches for assumption generation were developed further in~\cite{GP02,TDP03,Gupta:FMSD08,Nam:FMSD08}. Reference~\cite{TDP03} automatically generates the environment of Java programs from the specifications written by a user. \cite{Nam:FMSD08,Gupta:FMSD08} are concerned about automatic partitioning, learning, or minimizing assumptions for compositional verification. None of them considers environment generation for both
 model checking and testing.

Several specification-based environment generation methods exist: ~\cite{Parizek:EUROMICRO07} uses ADL to define protocols of Java components and constructs an environment
for the ADL specification. \cite{Dwyer:FSE98} describes environmental assumptions in LTL and uses them to filter a universal environment, which is adopted in
our approach to constrain the non-deterministic initial task model.
Reference~\cite{Yatake:SPIN10} automatically generates scripts in {\sc Promela} from environment models for OSEK/VDX-based operating systems that are modeled in {\sc Uml} diagrams.
Their approach, however, models all basic objects in OSEK/VDX using UML class diagrams and state diagrams, from which all combinations of deterministic environments
are generated and verified individually. The models are then used to automatically generate exhaustive test cases for the conformance testing of OSEK/VDX-compliant operating systems~\cite{Chen:APSEC11}.
Their approach assures the exhaustiveness of test cases, but the scalability issue remains, as the number of test cases may increase exponentially.

Program slicing~\cite{Weiser84} has been a popular technique for reducing verification complexity for both model checking and testing.
References~\cite{Binkley99} and \cite{Gupta92anapproach} use slicing algorithms to explicitly detect $def-use$ associations that are affected by a program change for efficient
regression testing. Reference~\cite{Chebaro:SANTE} performs program slicing for C programs with respect to the alarms generated from value analysis.
\cite{He:ICCD07} integrates aggressive program slicing and
a proof-based abstraction-refinement strategy for wireless cognitive radio systems. It is a representative example of using program slicing and bounded model checking for embedded software, but the slicing is integrated into the model checking process, and is thus not suitable for application in testing.


\section{Background}
\label{sec:background}
\subsection{OSEK/VDX}
{\sc Osek/vdx}
is
a joint project of the automotive industry,  which aims at establishing an industry standard for an open-ended architecture for distributed control units in vehicles~\cite{OSEK/VDX:Portal}.
The aim of {\sc Osek/vdx} is to provide standard interfaces independent of application, hardware, and network, and ultimately, to save the development costs for non-application
related aspects of control software.
It is specialized for automotive control systems, removing all undesired complexities such as dynamic memory allocation, circular waiting for resources,
multi-threading, and so on. Since its target system is safety-critical, it strictly prohibits uncontrolled dynamic behavior of the system.

Conformation testing is a standard verification method for the certification of OSEK/VDX-based operating systems. However, conformation testing suites
are typically insufficient to identify safety problems. As OSEK/VDX explicitly specifies more than 26 basic APIs, thorough conformation testing would require at
least $26\times 2\times 3$ test cases even if we assume two arguments per API and only boundary values for the arguments are chosen. The possible number of
execution sequences for these $26\times 2\times 3$ test cases would rise to 156 factorials, a large number to be tested in practice.

\subsection{Trampoline}

\begin{figure}
\centering
\includegraphics[width=108mm]{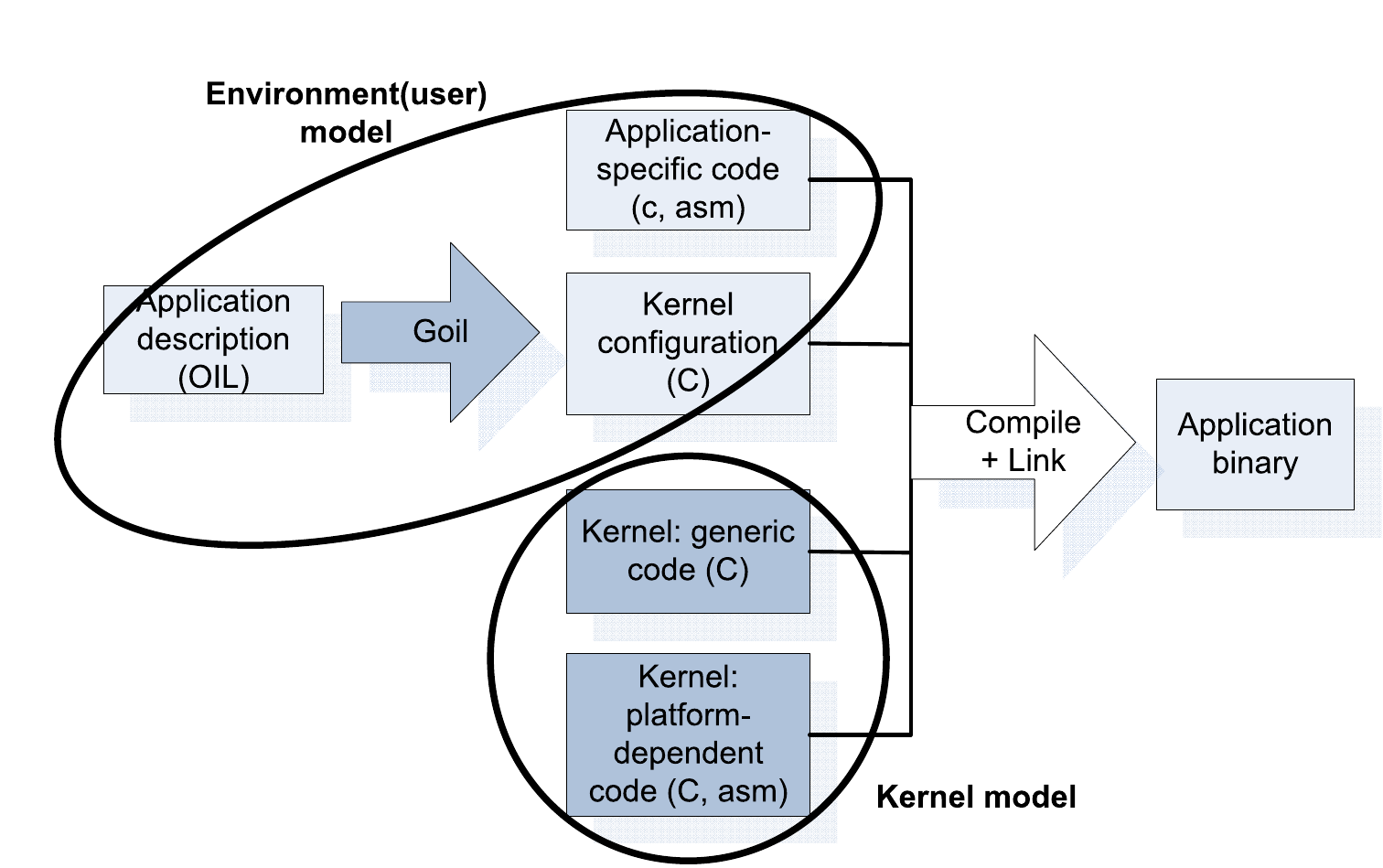}
\caption{Components of Trampoline\label{fig:trampoline}
}
\end{figure}

Trampoline~\cite{Trampoline:Portal} is an open source, real-time operating system compliant with OSEK/VDX version 2.2.3. It is developed in ANSI C and can be
ported to various hardware platforms such as Arm, POSIX, PPC, AVR, HCS12, C166, etc. Since it also supports POSIX, it can be test-run
on a UNIX/Linux environment before being ported to an actual operational environment.  As its target platform varies, its platform-dependent part is clearly
structured in a separate module that combines with the kernel module at compile time. Access to the hardware-specific part is abstracted using
$extern$ variables and macros so that the main control logic does not need to be aware of the specific hardware feature. As illustrated in Figure~\ref{fig:trampoline},
development of an automotive software using Trampoline requires four components; (1) application source code, (2) kernel configuration generated from
configuration description written in OIL (OSEK Implementation Language) using the Goil compiler, (3) generic OS kernel code compliant with the OSEK/VDX standard, and (4) platform-dependent
kernel code. The generic OS kernel code implements services for task management, resource management, interrupt handling, and event/counter/alarm management,
providing corresponding APIs.

\subsection{Model checking using CBMC}

Formal verification methods based on model checking~\cite{ModelCheckingBook} are an effective technique for identifying subtle issues in
software safety which is particulary important for embedded systems. Current technological advances in model checking enable engineers to
directly apply the technique to program source code, removing the manual model construction process. CBMC~\cite{CBMC:TACAS04} is one of these
model checking tools, which is capable of verifying almost full ANSI C. It can be used to verify buffer overflows, pointer safety, exceptions
and user-specified assertions. Furthermore, it can check ANSI C and C++ for consistency with other languages, such as Verilog. The main
advantage is that it is completely automated and generates counterexample traces when a property in question is refuted.

As with any other model checking tool, CBMC also suffers from the problem of scalability. When applied to the Trampoline kernel as a whole with
an arbitrary sequence of API calls, for example, it ran
out of memory for checking one assertion on a PC with 3GB of memory.

\section{Overall Approach}
\label{sec:overview}

Comprehensive verification, required by functional safety analysis, is too costly to be applied in practice. Reducing the cost while maintaining
comprehensiveness is a challenging, but crucial task.
Our approach attempts to achieve this goal with the following three strategies:

\begin{enumerate}
\item
Property-based environment generation: An environment of the operating system kernel is automatically generated using static code analysis for a given
     safety property.
\item
Property-based abstraction: The operating system kernel is abstracted by extracting only the code relevant to a given property.
\item
Collaborative verification using model checking and testing: Both model checking and testing are used complementarily for the verification of
 the abstract kernel code under the generated environment model.
\end{enumerate}

\begin{figure}
\centering
\includegraphics[width=132mm]{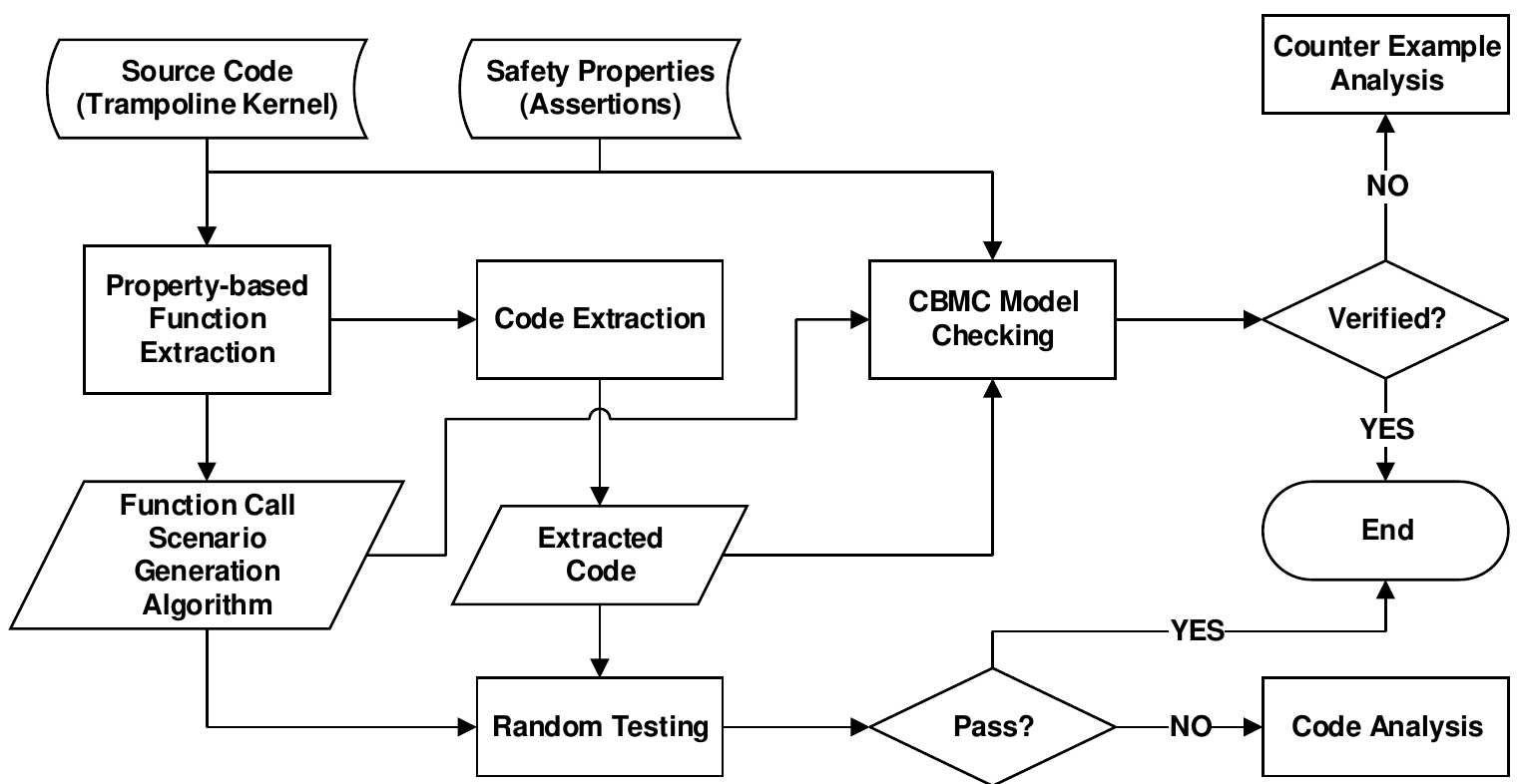}
\caption{Collaborative verification approach\label{fig:approach}
}
\end{figure}

Cost reduction is achieved through property-based environment generation and code abstraction. The efficiency of verification is increased
by taking advantage of both verification techniques. Figure~\ref{fig:approach} is an overview of the suggested collaborative verification
approach. Our approach uses both model checking and testing to complimentarily utilize their different capabilities when only limited resources are available.

\section{Environment Generation}
\label{sec:env_gen}

A straightforward way to include all possible task interactions with the operating system is to model the task with strongly connected states, where each state represents an API call to the kernel and each transition between states is not guarded.
However, this includes too many spurious and/or impossible behaviors and
increases the cost for verification as well as counterexample analysis; if 26 APIs are provided by the operating system, the task model would have at least 26 strongly connected states.
Our approach tries to minimize unnecessary verification cost by using property-based extraction of dependent functions.

\subsection{Abstraction through static code analysis}
Given a property, we first extract the variables specified in the property, which is called {\it Verification Target Variables}, and identify
all the variables that are used to define the {\it Verification Target Variable}, called {\it Extended Verification Target Variable}.
Then, functions modifying those {\it Extended Verification Target Variables},
called $End\_Level\_Functions$, are extracted.
The prototypes of the $End\_Level\_Functions$ are used to construct an end-level environment model.
The corresponding end-level abstract kernel code consists of all the $End\_Level\_Functions$ and their dependent code.
The {\it Root\_Level\_Functions} are identified by performing backward reachability analysis from each {\it End\_Level\_Function}.
The prototypes of the $Root\_Level\_Functions$ are used to construct a root-level environment model. Its corresponding abstract kernel
code is identified by performing forward reachability analysis from each $Root\_Level\_Function$. The result of forward reachability analysis
is also used to identify constraints for the end-level environment model.

\begin{definedef}
Property-related variables:
\begin{enumerate}
\item
A Verification Target Variable is a variable that appears in the property specification.
\item
An Extended Verification Target Variable is a variable that a Verification Target Variable depends on.
\end{enumerate}
\end{definedef}

\begin{definedef}
Classification of functions:
\begin{enumerate}
\item An {\it End\_Level\_Function} is a function that directly modifies, sets, or uses an Extended Verification Target Variable.
\item A {\it Root\_Level\_Function} is an API that is a terminal node of the called-by graph of an {\it End\_Level\_Function}.
\end{enumerate}
\end{definedef}

\begin{figure}
\centering
\includegraphics[width=156mm]{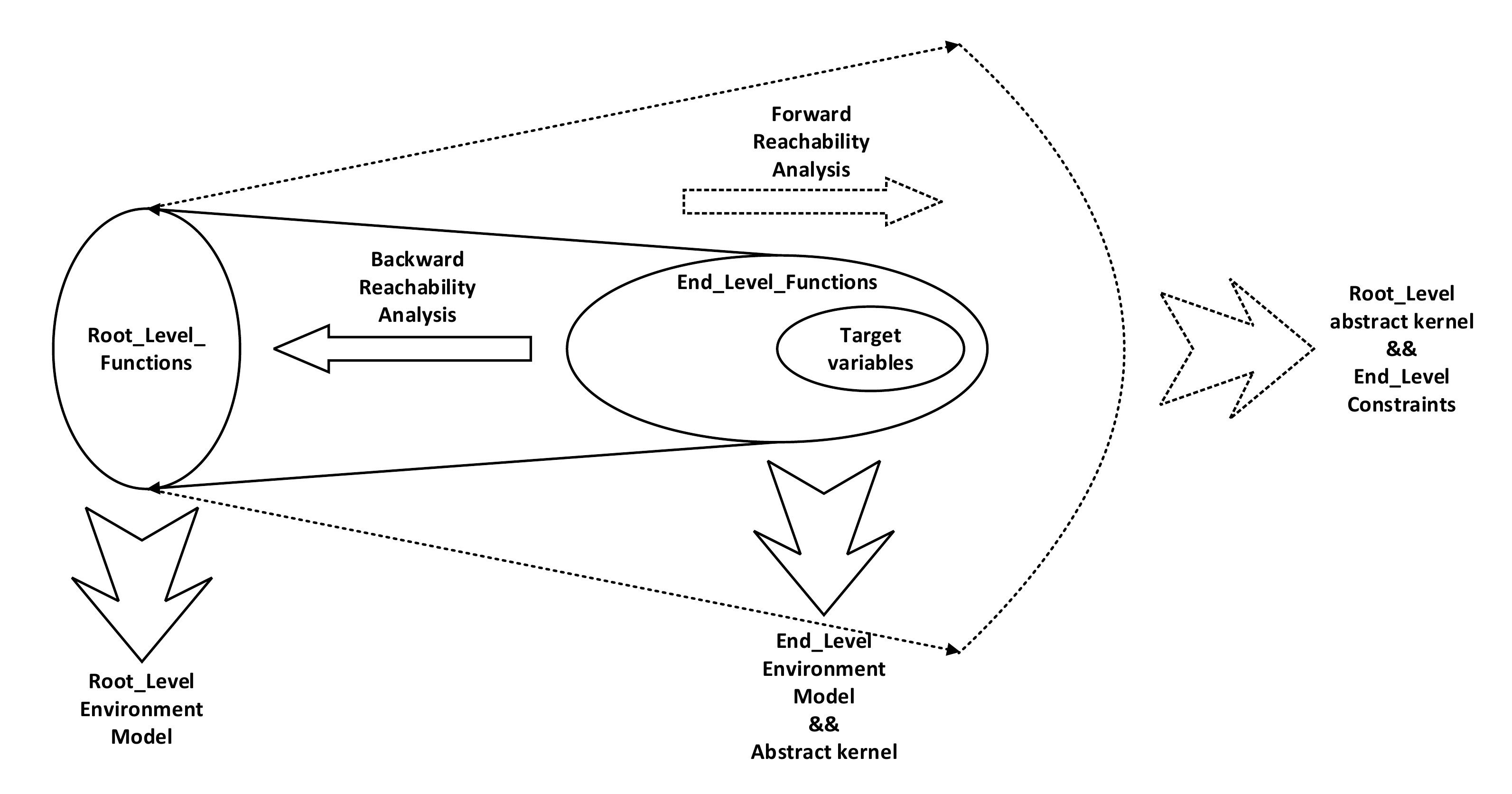}
\caption{Backward and forward reachability analysis for environment generation \label{fig:env_gen_concept}
}
\end{figure}

From this process, we extract two types of functions for constructing different levels of  environment models:
(1) functions for root-level environments,  and (2) functions for end-level environments.
Figure~\ref{fig:env_gen_concept} shows the conceptual diagram for the whole process.

For a simple example, if a property in question is
\[ Property_1: \ assert(tpl\_fifo\_rw[tpl\_h\_prio].size > 0),\]
then we first identify {\it Extended Verification Target Variables} and {\it End\_Level\_Functions} for {\it tpl\_fifo\_rw} and {\it tpl\_h\_prio}.
The identified set of $End\_Level\_Functions$ for the variable $tpl\_h\_prio$ is
\{ $tpl\_get\_proc$, $tpl\_put\_preempted\_proc$,
$tpl\_put\_new\_proc$, $tpl\_schedule\_from\_running$ \} in the Trampoline kernel.
An end-level environment model is constructed as non-deterministic calls to those end-level functions and
its corresponding abstract kernel encompasses all the identified $End\_Level\_Functions$ and their dependent code.
We then identify {\it Root\_level\_Functions} for each of the {\it End\_Level\_Functions}; For examples, \{$ReleaseResource$, $Schedule$, $ActivateTask$, $SetEvent$, $TerminateTask$,
$ChainTask$, $WaitEvent$, $StartOS$\} are {\it Root\_Level\_Functions} for $tpl\_get\_proc$, which are identified from its called-by graph.
A root-level environment model consists of non-deterministic calls to those API functions and its corresponding
abstract kernel encompasses all the identified root level functions and their dependent code.

\subsection{Implementation}

\begin{figure}[h]
\centering
\includegraphics[width=156mm]{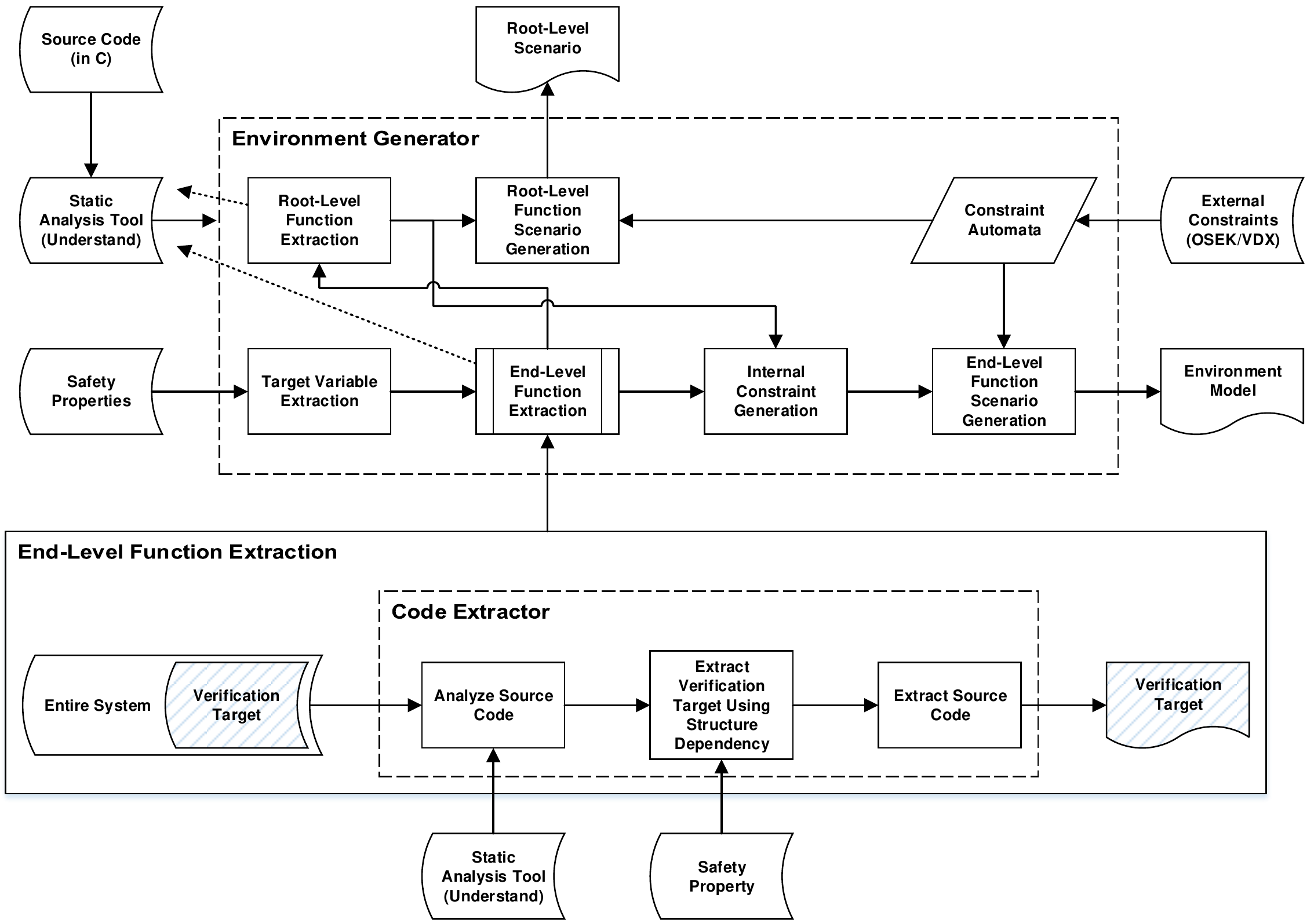}
\caption{Environment Generation \label{fig:env_gen}
}
\end{figure}
The suggested approach was implemented and fully automated. Figure~\ref{fig:env_gen} shows the overall structure of the automation.
The source code of the Trampoline OS is analyzed by the static analysis tool Understand~\cite{Understand}, which creates a data repository of the analysis results from which information on variable/function dependencies can be extracted using a C plug-in.
The environment generator first extracts target variables from the properties and then extracts Extended Target Variables and End-Level Functions by analyzing dependency relations among variables and functions. Root-Level Functions are then extracted from the called-by graph for each End-Level Function. A Root-Level Function Scenario is generated as an arbitrary sequence of function calls of Root-Level Functions, which complies with the external constraints from the OSEK/VDX standard.  Finally, the environment model is generated as an arbitrary sequence of End-Level Functions that complies with both the external and the internal constraints. The internal constraints are partial-order relations among End-Level Functions, which are generated from each Root-Level Function.

The last step is the property-based abstraction of the original code. Since the environment generation step identifies all necessary End-Level Functions modifying the Verification Target Variables together with the ordering relation among them in call sequences, verification requires only the source code of those End-Level Functions plus codes for preprocessing them. Therefore, for each safety property, verification is performed using the environment model generated with the tool and the property-based abstract code.

\section{Setting up Environments for Collaborative Verification}
\label{sec:env_setting}

\begin{figure}[h]
\centering
\includegraphics[width=132mm]{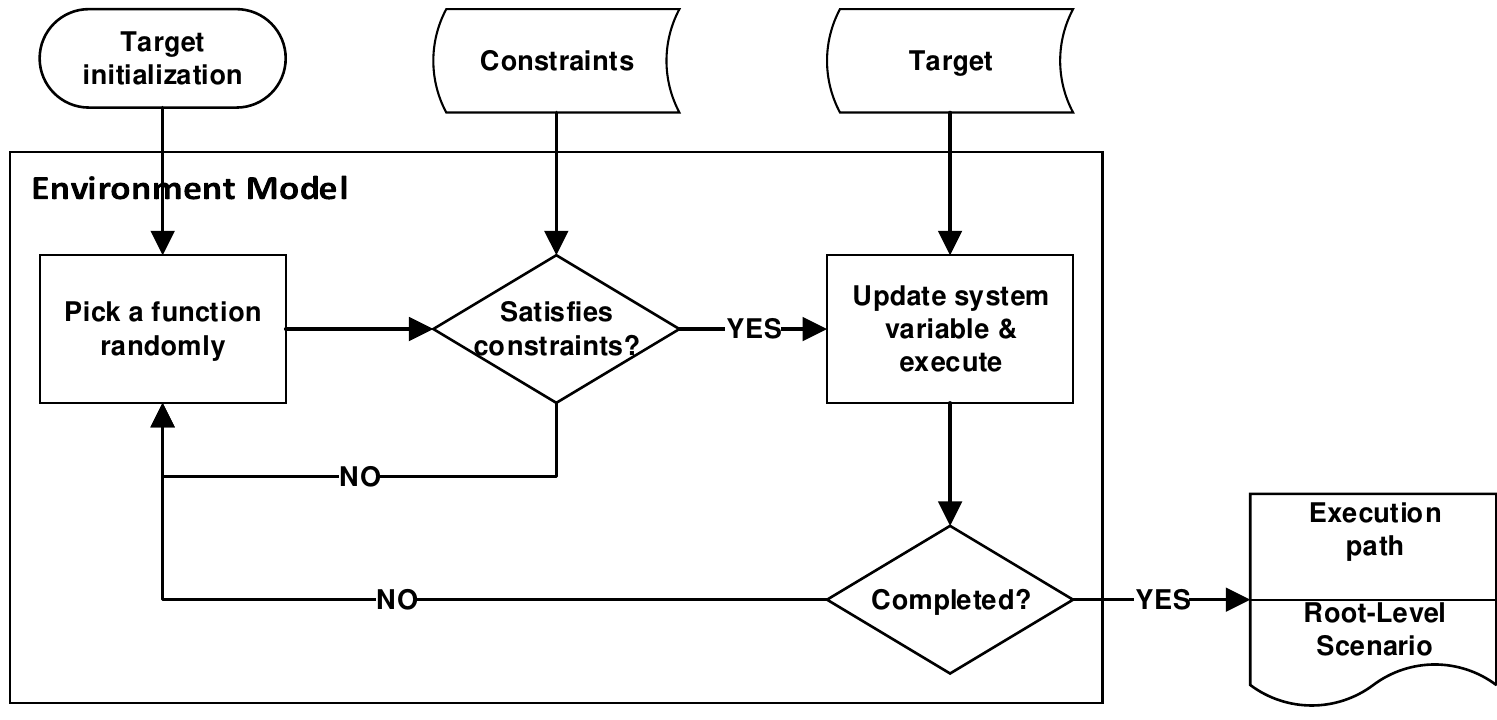}
\caption{Scenario generation process for random testing\label{fig:envModel}
}
\end{figure}

CBMC and testing require different settings for their verification environments even with the same set of $End\_Level\_Functions$; CBMC requires only the algorithm of non-deterministic function calls.
Random testing (both root-level and end-level), however, requires explicit function call sequences generated from a given environment model.
Figure~\ref{fig:envModel} shows the process for generating such function call sequences, which can be applied to both End-Level and Root-Level environment models.
It repeats the selection and checking process by arbitrarily selecting a function and checking whether the selected function satisfied the constraints or not.
We have implemented an OSEK/VDX simulator to check the external constraints in the process of root-level sequence generation.  Internal constraints are identified by call graph analysis.
Details are described in the following two sub-sections.

We note that the checking the constraints is not necessary to ensure the correctness of the scenario generation, but it is essential for making the verification efficient, since otherwise the verification produces too many false errors.

\subsection{Root-Level Scenario Generation}

\begin{figure}[h]
\centering
\resizebox{\linewidth}{!}{
\includegraphics[width=80mm]{trampoline_fig.pdf}
\includegraphics[width=70mm]{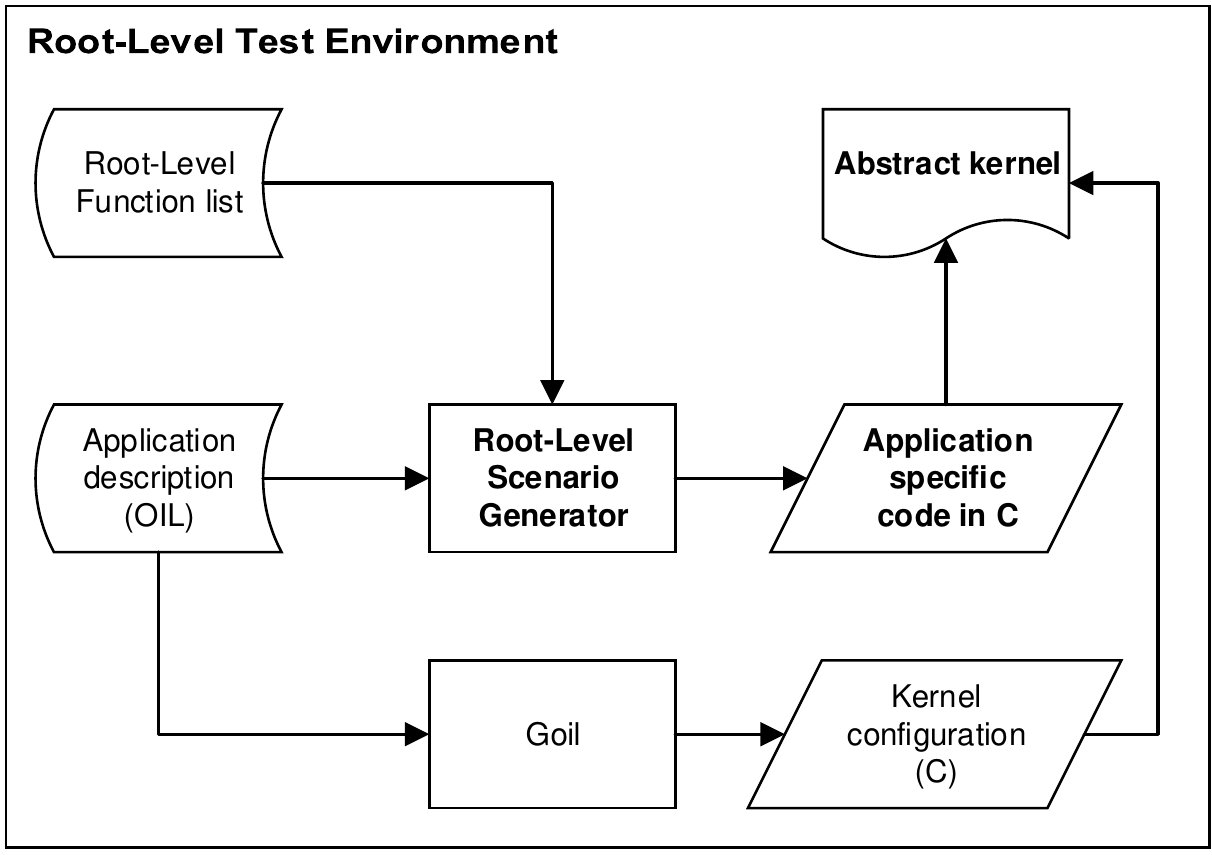}
}
\caption{Root-Level scenario and the testing environment\label{fig:RootLevelScenarioGeneration}
}
\end{figure}

Root-Level scenario generation is based on the scenario generation process illustrated in Figure~\ref{fig:envModel}. A Root-Level Function Scenario is an arbitrary sequence of function calls of Root-Level Functions, and it complies with the external constraints from the OSEK/VDX standard. Figure~\ref{fig:RootLevelScenarioGeneration} shows the Root-Level test environment for the Trampoline OS. Since an OIL file is required for testing, an OIL file is specified as an input, which is compiled with the extracted kernel code and the random sequence of Root-Level Functions generated from our Root-Level scenario generator.

There are two things to be done before implementing a scenario generator. First, every constraint specified in the OSEK/VDX standard should be identified. Second, an OSEK/VDX Simulator - an abstract OSEK/VDX model - should be implemented in order to trace all the changes and fully observe constraints.

\subsubsection{Identification of external constraints}

\begin{figure}[h]
\centering
\includegraphics[width=156mm]{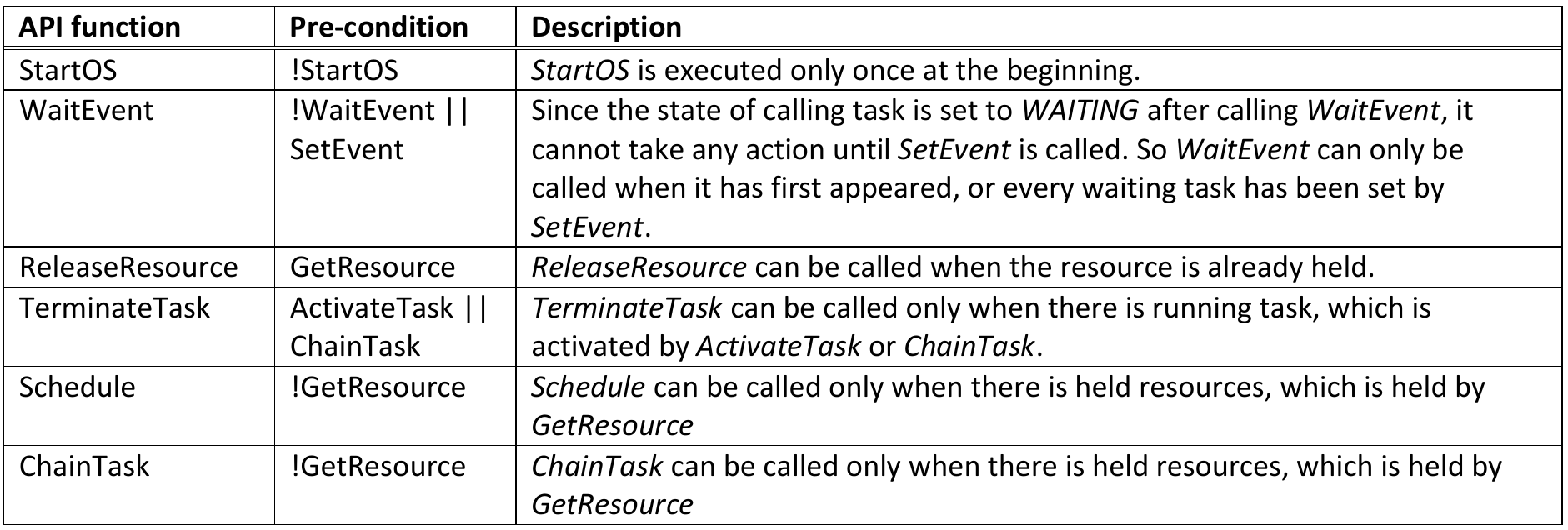}
\caption{Constrant list extrected from OSEK/VDX spec\label{fig:constraint}
}
\end{figure}

The OSEK/VDX standard explicitly specifies constraints among the APIs. The $description$ column of Figure~\ref{fig:constraint} lists some of the constraints manually identified
from the standard. These constraints are represented as pre-conditions with respect to other APIs. For example, the API function $TerminateTask$ can be called only if the task has been activated either
by $ActivateTask$ or $ChainTask$. Therefore, we set \{$ActivateTask$, $ChainTask$\} as preconditions of $TerminateTask$. Figure~\ref{fig:constraint} shows a couple of preconditions of other API functions.

\begin{figure}[h]
\centering
\includegraphics[width=137.5mm]{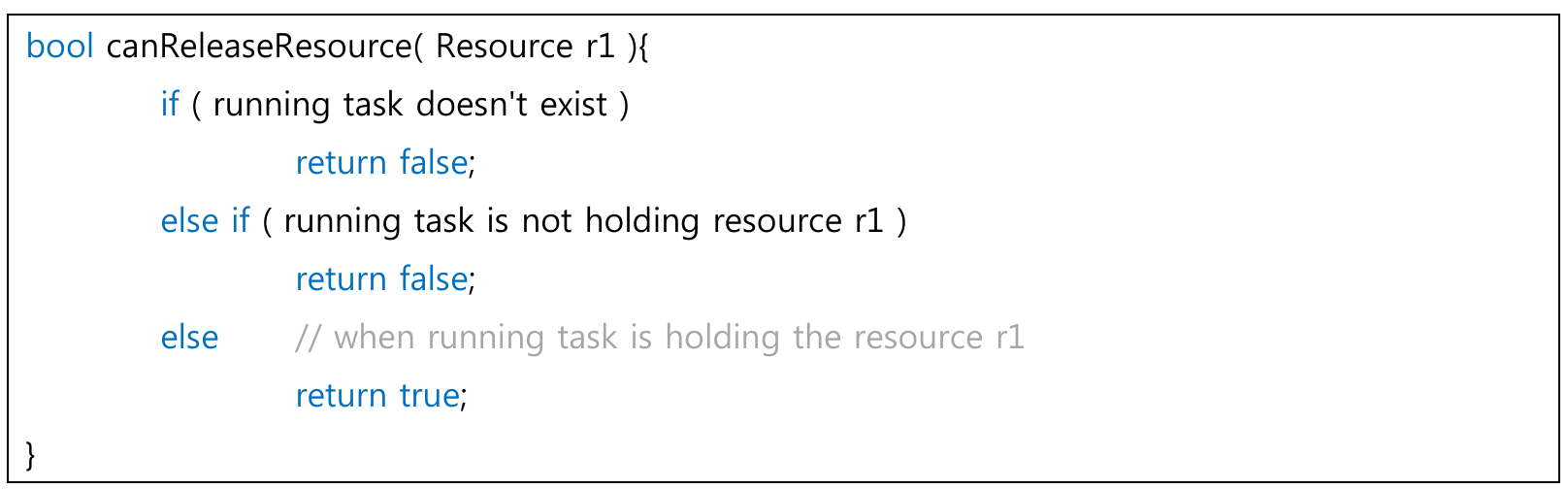}
\caption{ReleaseResource constraint checker\label{fig:constraintCheck}
}
\end{figure}

Identified constraints are then imposed in the Root-Level scenario generation process illustrated in Figure~\ref{fig:envModel}.
Figure~\ref{fig:constraintCheck} is an example algorithm of the constraint checking.

\subsubsection{OSEK/VDX Simulator}
To fully consider all the identified constraints, it is necessary to trace changes that previous function calls have made. For example, if {\it ActivateTask(t1)} is chosen as the first Root-Level Function in a scenario, task {\it t1} should be marked as $READY$ task, for further scenario validation. This process is fully automated by implementing an OSEK/VDX simulator.

\begin{figure}[h]
\centering
\includegraphics[width=132mm]{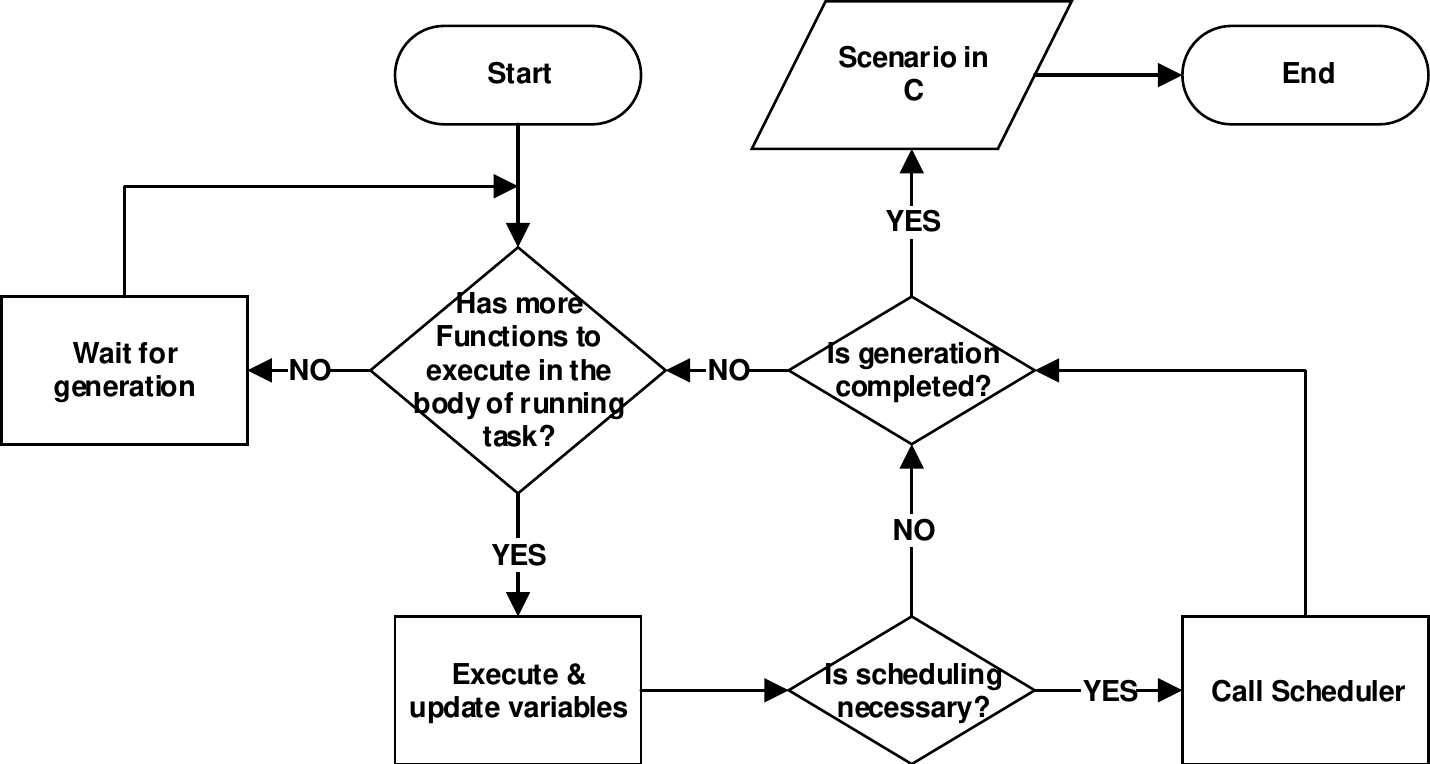}
\caption{OSEK/VDX simulator\label{fig:simulator}
}
\end{figure}

The OSEK/VDX simulator traces run-time information such as list of resources, list of events, list of task models, reference to running task, ready queue (priority queue), and waiting queue. It provides Root-Level Function calls just like OSEK/VDX APIs. When one of these procedures is called, it simulates the behavior of OSEK/VDX. The simulator includes task model, task scheduler, event management, and resource management. Figure~\ref{fig:simulator} shows the overall process of the OSEK/VDX simulator.

A randomly chosen function is added to the body of the running task by a module called Scenario Generator. When there is a function in the body of the running task that has not been executed by the simulator, it executes the function. If scheduling is necessary, it calls the scheduler and executes all the functions that already exist in the body of the preempted task, when preemption occurs. This process is repeated until there is no function left to execute, and the Scenario Generator is requested to generate another function randomly.

Whether scenario generation has been completed or not can be determined by checking all tasks that are initially generated from the given OIL file. If every task has completed its execution with $TerminateTask$ or $ChainTask$, it can be determined that the scenario generation is completed.

\subsection{End-Level Environment Model}
The End-Level scenario generation process is also based on the scenario generation process illustrated in Figure~\ref{fig:envModel}.
It executes an arbitrary sequence of $End\_Level\_Functions$, which is chosen from among the list of $End\_Level\_Functions$ serving as an environment of CBMC model checking and End-Level random testing.

This process considers internal constraints among End-Level functions; since End-Level functions can only be called by API-Level functions, the pre-conditional constraints of Root-Level functions identified from the OSEK/VDX constraints must be implicitly obeyed by the End-Level functions. These implicit constraints can be identified by analyzing call-graphs of Root-Level functions and their pre-conditional
relations. We call such implicit constraints internal constraints.

\begin{figure}[h]
\centering
\includegraphics[width=100mm]{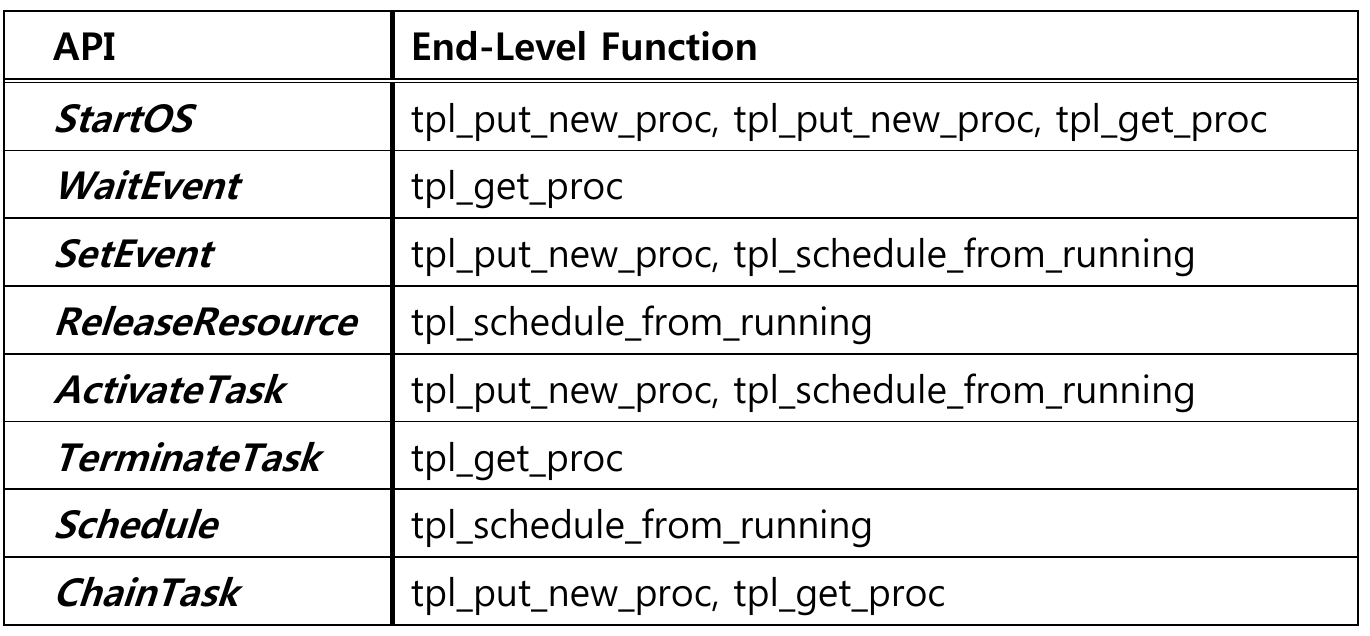}
\caption{End-Level Functions called by each API-Level Function\label{fig:callgraph}
}
\end{figure}

For example, if we consider the Root-Level APIs and their corresponding End-Level functions, the external constraint {\it $WaitEvent$ can be called after SetEvent is called} can be re-interpreted as {\it $tpl\_get\_proc$ can be called after $tpl\_put\_new\_proc$ and $tpl\_schedule\_from\_running$ are called}, which can be written in a regular expression $(tpl\_put\_new\_proc\ tpl\_schedule\_from\_running\ (End\_Level\_Function\ -\ tpl\_get\_proc)^*\ tpl\_get\_proc)^*$.

Checking constraints of this kind cannot be done using the OSEK/VDX simulator because the End-Level functions include implementation-specific function names that cannot be modeled from the standard.
Instead, the internal constraints are simplified using the characteristic that {\it a function cannot be called more times than its preceding functions in the partial order relation}. The example internal constraint can be simplified as {\it The number of $tpl\_get\_proc$ calls cannot exceed the one of either $tpl\_put\_new\_proc$ or $tpl\_schedule\_from\_running$}. The constraint checker keeps track of the
number of each End-Level functions calls and checks $(\#\ tpl\_get\_proc < \# \ tpl\_put\_new\_proc)\ \&\&\ (\#\ tpl\_get\_proc < \#\ tpl\_schedule\_from\_running)$.

\section{Experiments}
\label{sec:experiment}

We have conducted a series of experiments to show the impact of our approach using CBMC model checking,  End-Level random testing, and Root-Level random testing.
The target verification properties are three functional safety properties from the Trampoline kernel:

\begin{eqnarray}
& &assert(tpl\_h\_prio \ne -1) \\
& &assert(tpl\_kern \ne NULL) \\
& &assert(tpl\_kern\rightarrow state == RUNNING)
\end{eqnarray}

$tpl\_h\_prio$ is the value of the highest-priority task in the ready queue in the Trampoline kernel.
$tpl\_h\_prio \ne -1$ is supposed to be true whenever rescheduling is necessary.
$tpl\_kern$ stores the key information of the currently running task.
$tpl\_kern \ne NULL$ and $tpl\_kern\rightarrow state == RUNNING$ checks if the state of the running task is $RUNNING$ when the scheduler is called.

We performed verification of these three assertions using the model checker CBMC, Root-Level random testing, and End-Level random testing.
The verification cost in terms of the number of verification conditions and the resource requirements was measured for CBMC verification.
Branch coverage was measured using the Squish Coco code coverage tool~\cite{SquishCoco} for random testing.
All experiments were performed  on Linux Fedora 16 OS, with Intel Xeon 3.4GHz e3-1270 processor and 32GB of 1333MHz DDR3 RAM.

\begin{figure}[h]
\centering
\includegraphics[width=154mm]{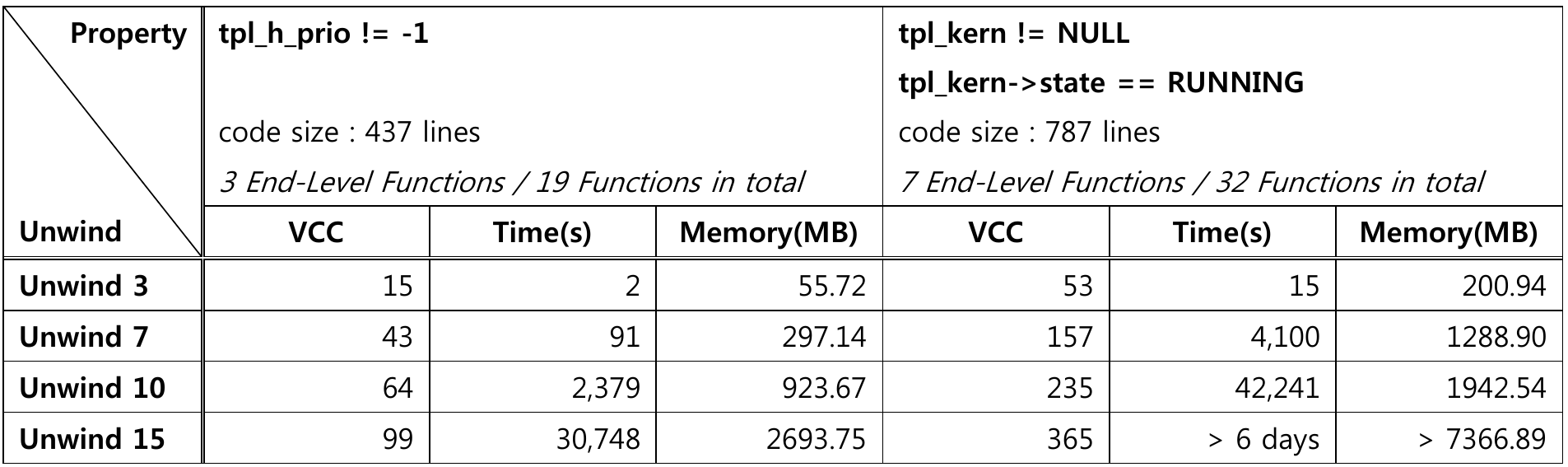}
\caption{Time and memory space to verify with CBMC\label{fig:table_mc}
}
\end{figure}

Figure~\ref{fig:table_mc} shows the time and memory space it took to verify the Trampoline operating system with the End-Level Environment Model using the model checker CBMC. Time and memory space increase exponentially as
the length of the End-Level function calls (unwind value) increases. CBMC verifies the assertions by searching through every possible scenario with the length of the unwind value, making it a powerful method. With the given resources, CBMC reported no counter examples up to the unwind value 10, but was not able to finish its verification process for the unwind value 15 after 6 days.
We do not report the result of CBMC model checking using Root-Level environment models since it is too costly to perform even with the value of unwind option 10.

\begin{figure}[h]
\centering
\includegraphics[width=\linewidth]{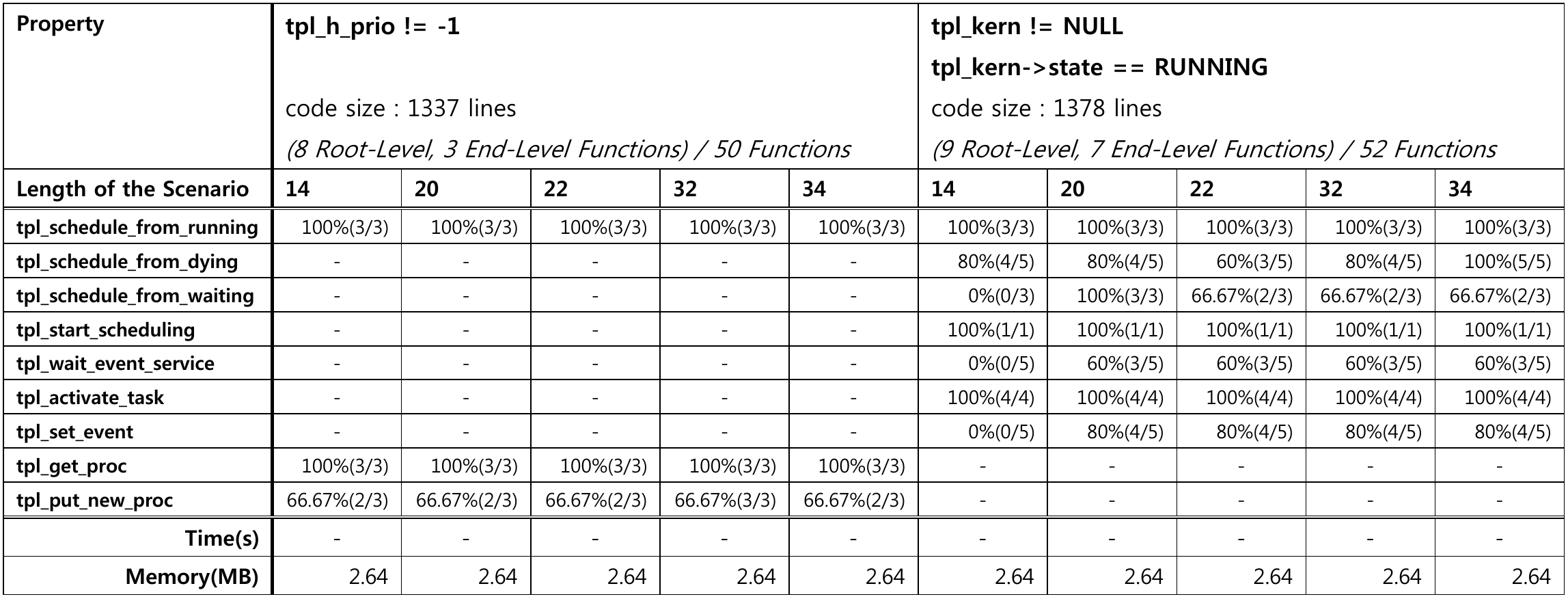}
\caption{Coverage, time, and memory space to verify with Root-Level Random Testing\label{fig:table_apiLevel}
}
\end{figure}
\begin{figure}[h]
\centering
\includegraphics[width=\linewidth]{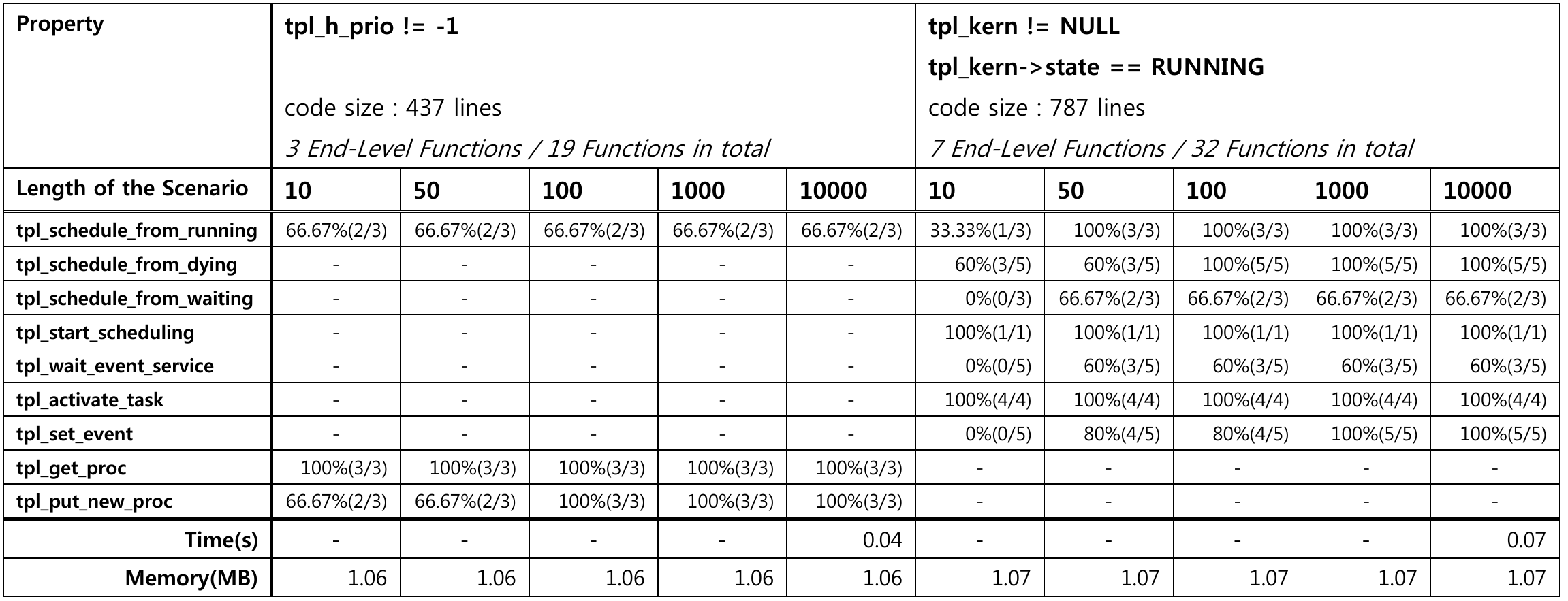}
\caption{Coverage, time, and memory space to verify with End-Level Random Testing\label{fig:table_endLevel}
}
\end{figure}

There are a few dozens of extracted functions in random testing environments, including End-Level Functions, as illustrated in Figures~\ref{fig:table_apiLevel} and ~\ref{fig:table_endLevel}. Due to a lack of space, only the test results of End-Level Functions are described. Root-Level Random Testing (Figure~\ref{fig:table_apiLevel}) is much faster (less than 1/100 seconds), consumes little memory (up to 2.64MB of memory), and achieves a certain level of test coverage quickly, but the coverage does not improve after test sequences of length 34.  In End-Level Random Testing (Figure~\ref{fig:table_endLevel}), a test sequence of length 100 achieves a certain level of  coverage both for $tpl\_h\_prio$ and $tpl\_kern$. The coverage stays the same afterwards. End-Level Random Testing required around 1.06~1.07 MBytes of memory.

For many cases, the coverage did not increase even with lengthier test cases. There can be two reasons why some part of the code are unreachable. The first reason is exception handling; parts of the code for exception handling are never reached unless an exceptional situation occurs. The second reason is that some variables in conditional statements are not included in the Extended Verification Target Variable.
So the behavior of updating these variables might not be fully extracted, which can make some conditional statements fixed.
\begin{figure}[h]
\centering
\includegraphics[width=120mm]{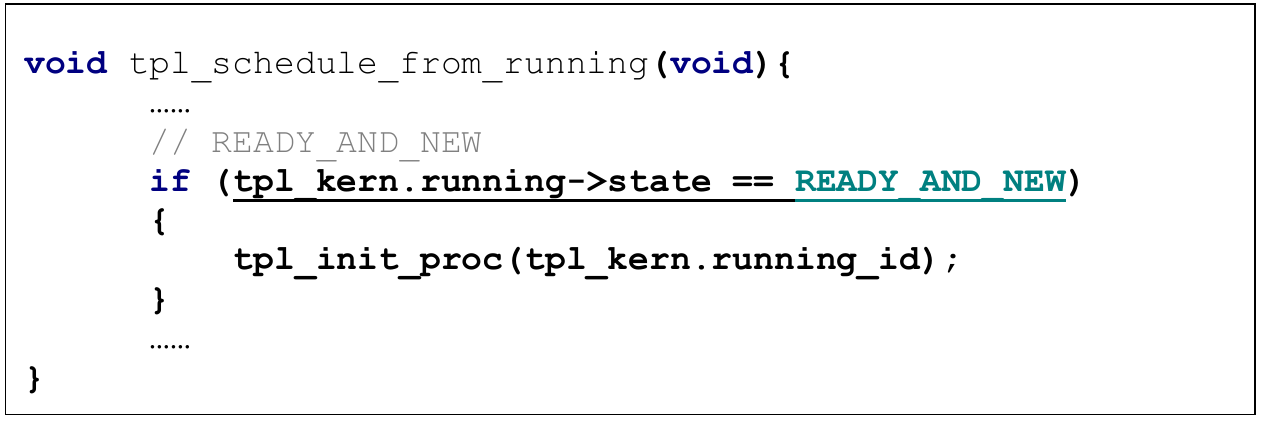}
\caption{Example of uncovered conditional statement\label{fig:code_uncovered}
}
\end{figure}
An example of this case is illustrated in Figure~\ref{fig:code_uncovered}. This conditional statement is only executed when {\it tpl\_schedule\_from\_dying} or {\it tpl\_activate\_task} is executed before this code. But these two functions are out of boundary in this model because the model is generated only with regard to the property {\it “tpl\_h\_prio != 0”}. Thus coverage cannot be increased, and testing terminates.

In terms of comprehensiveness, CBMC is the most powerful method, verifying every possible scenario within the same length, but it is limited by the length. As shown in Figure~\ref{fig:table_mc}, the verification cost increases exponentially as unwinding depth increases. Therefore, CBMC cannot detect potential faults that can be identified only in long task scenarios.
Unlike verification with CBMC, the length of the scenarios is not limited in random testing since the cost is much cheaper than CBMC, as shown in Figure~\ref{fig:table_endLevel}.
Though it cannot be comprehensive, it can be more effective in stress testing, since the length of the test sequences can be sufficiently long.

\begin{figure}
\centering
\includegraphics[width=\linewidth]{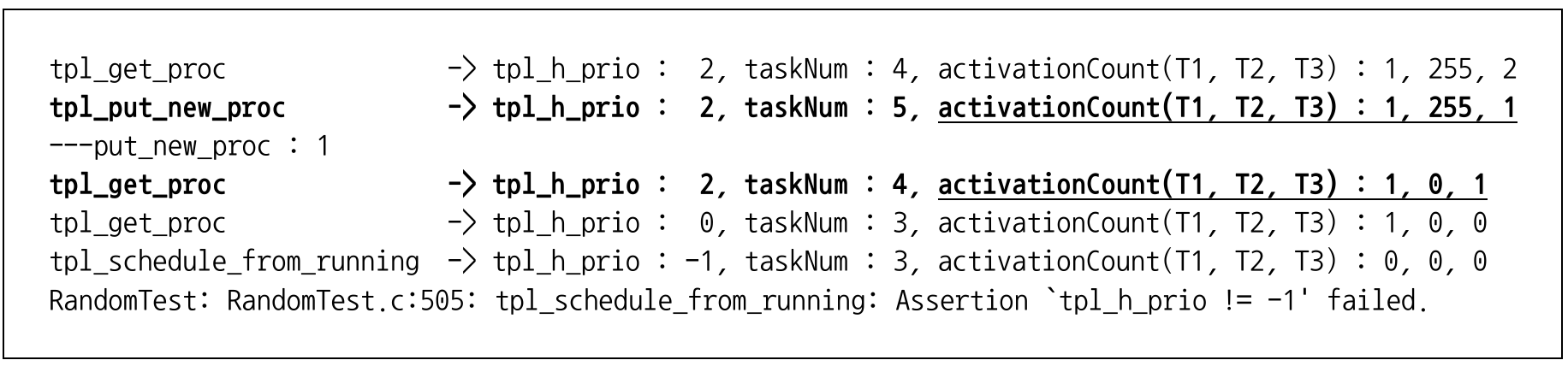}
\caption{Error caused by overflow\label{fig:overflow}
}
\end{figure}

End-Level Random Testing did in fact, catch some overflow errors in the Trampoline kernel as illustrated in Figure~\ref{fig:overflow}.
These errors have occurred because the size of the variables saving the activation count is limited to 8 bits;
the second line of Figure~\ref{fig:overflow} shows that $Task2$ has the activation count 255, but adding another activation changed its value to 0.
So Trampoline has changed the value of $tpl\_h\_prio$ to -1, meaning that the process table has no activated task available.
The variable size is implementation-specific and is not constrained to 8 bits, neither in the OIL specification nor in the OSEK/VDX specification. This problem could be addressed by constraining the size to 8 bits in the OIL specification.

A Model checker could find this type of potential faults if we can set the value of the unwind option larger than 255, but our experiments could not identify it due to resource limitations.
 End-Level Random Testing is appropriate for finding this kind of errors because the cost does not increase much even with lengthy test scenarios.

We could not identify the same fault using Root-Level Random Testing, either.
The main reason is that Root-Level Random Testing is coupled with a pre-defined OIL configuration file. The OIL file specifies the typical system configuration, and thus, activating a task over 255 times
is not likely to happen unless we specifically aim at stress testing. End-Level Random Testing is more effective in stress testing, since it is not constrained by the system configuration and can test abnormal cases.

API-Level Random Testing, however, is beneficial in that it is not necessary to do additional API-Level analysis when testing identifies faults, which is necessary in model checking and End-Level Random Testing.

\section{Conclusion}
\label{sec:conclusion}

This paper presented methods and tools for environment generation and code abstraction to improve the efficiency of verification using model checking and testing.
The effect of using the suggested approach was demonstrated through a series of experiments using the Trampoline operating system as a case example.
The benefit of property-based environment generation is two-fold: (1) it reduces verification cost by reducing the target code and by limiting its environment to the task interaction
scenario relevant to the verification property, and (2) it simplifies the analysis process and localizes the verification activity by focusing on the points of interest.

The experiments revealed relative pros and cons of the three verification methods and identified potential safety faults, which suggests the following collaborative use of model checking and testing;
\begin{enumerate}
\item
Apply End-Level Random Testing first for stress testing.
\item
Apply Root-Level Random Testing to conform the errors identified through End-Level Random Testing.
\item
Apply model checking using CBMC last for comprehensive verification within a limited scope.
\end{enumerate}

Our tool still needs some improvements. First, conditional dependencies need to be considered so that test coverage can be improved.
Second, Root-Level scenario generation currently assumes a fixed OIL configuration. We would like to relax the condition so that an arbitrary OIL can be handled by the tool.

\bibliographystyle{eptcs}
\bibliography{E:/research/papers/bib/cbe_mod}
\end{document}